# Longitudinal NMR and Spin States in the A-like Phase of ³He in Aerogel

V.V. Dmitriev[1], L.V. Levitin[1], N. Mulders[2], D.E. Zmeev[1]

[1]*Kapitza Institute for Physical Problems, 2 Kosygina Str., Moscow 119334, Russia*
[2]*Department of Physics and Astronomy, University of Delaware, Newark, Delaware 19716, USA*

**Abstract.** It was found that two different spin states of the A-like phase can be obtained in aerogel sample. In one of these states we have observed the signal of the longitudinal NMR, while in another state no trace of such a signal was found. The states also have different properties in transverse NMR experiments. Longitudinal NMR signal was also observed in the B-like phase of ³He in aerogel.

## INTRODUCTION

An aerogel is a "tangle" consisting of $SiO_2$ strands. Diameter of strands is about 30 angstroms i.e. much less than the correlation length in superfluid ³He. Consequently the system "liquid ³He + aerogel" allows to investigate an influence of impurities on superfluidity with nontrivial Cooper pairing. It is known that in a weak magnetic field two superfluid phases (called A-like and B-like) can exist in ³He in aerogel [1,2]. It is also established [1,3] that the B-like phase is analogous to the B-phase of "usual" bulk ³He (i.e. is described by similar order parameter). However the problem of identification of the A-like phase remains unsolved. At pressures above 20 bar this phase can exist in a wide range of temperatures below the superfluid transition temperature $T_{ca}$ in a metastable (supercooled) state. It is established [1] that the A-like phase belongs to the family of Equal Spin Pairing (ESP) phases, but its NMR properties do not correspond to the properties of the A-phase of bulk ³He [1,4-6]. I.A.Fomin has suggested the so called "robust" ESP phase – the phase in which the orientation of the order parameter is not influenced by presence of the aerogel – as a possible candidate for the A-like phase [7,8]. G.E.Volovik suggests another state as a model of the A-like phase – the state with highly spatially randomized bulk A-phase order parameter [9,10]. D.D.Osheroff [11] based on analysis of some experimentally measured values suggested that the A-like phase is a planar (two-dimensional) phase.

We present here results of our recent NMR studies of the A-like phase which may be useful for its identification. One of the aims of our experiments was an observation of longitudinal NMR signal from the A-like phase of ³He in aerogel and comparison of the results with transverse NMR data.

## EXPERIMENTAL SETUP

Experiments were done at pressures 26.0 bar and 29.3 bar in magnetic field of 224 Oe (corresponding to NMR frequency of 728 kHz). We used so called 98.2% porosity aerogel, in which 98.2% of the volume is empty and silica strands occupy only about 1.8% of the whole volume. Note that in the most of experiments done by other groups, aerogel samples with the same (or close) porosity were used. Our aerogel sample had a cylindrical form (diameter=4mm, length=6 mm) with the axis oriented along the external steady magnetic field. The sample was situated inside the epoxy cell, so that there were only small gaps (about 0.07 mm) between the sample and the cell walls (Fig.1). The cell was surrounded by transverse (60 turns) and longitudinal (350 turns) NMR coils. Both coils were wound from thin (diameter of 0.07 mm) superconducting NbTi wire in CuNi clad. During the first cooldown, in order to avoid disturbances of the external magnetic field homogeneity, both coils were cooled through the superconducting transition temperature (~9 K) in the presence of the NMR field (see Ref.12 for details of the procedure). As a result the width of the $^3$He transverse NMR line in the normal phase was usually less than 25-30 Hz. Standard NMR setups were used, i.e. radiofrequency (RF) excitation was applied to the NMR circuit; the voltage across the corresponding coil was amplified by a preamplifier and then detected by a lock-in amplifier. In both NMR spectrometers we have used room temperature preamplifiers but the NMR circuit for longitudinal resonance was cold and had the fixed frequency (9593 Hz). The quality factor of this circuit was about 2300.

The temperature was obtained by copper nuclear demagnetization refrigerator and was measured with a vibrating wire situated in a large volume connected with the cell by narrow (diameter of 1 mm) channel. Two heaters were used to stabilize and to sweep the temperature.

In order to avoid the paramagnetic signal from solid $^3$He on the surface of aerogel strands, the aerogel sample was preplated by about 2.5 monolayers of $^4$He. Consequently no Curie-Weiss behavior of spin susceptibility was observed in our experiments.

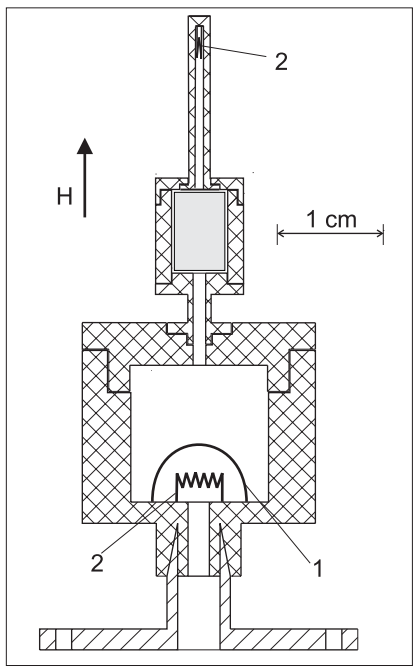

**FIGURE 1.** The experimental cell. Aerogel sample is shown in gray. 1 – vibrating wire, 2 – heaters.

## LONGITUDINAL NMR EXPERIMENTS

The experiments on longitudinal NMR were performed in the following way. Resonant frequency excitation (usually corresponding to the amplitude of the RF field of 0.02-0.1 Oe) was applied to the longitudinal NMR circuit. Then we monitored lock-in output signals and started to change the temperature by changing the power, dissipated by heaters (Pos. 2 in Fig.1). The results of such an experiment are shown in Fig.2. Blue curves correspond to cooling from above $T_{ca}$ down to temperatures below the temperature of A-like → B-like phase transition ($T_{ab}$). Here we at first get the A-like phase and then at $T= T_{ab}$ transition into B-like phase takes place. Note that this transition is rather slow (it occurs during about 1 minute) but the corresponding change in the longitudinal signal is very clear. Red curves correspond to a subsequent warming from the B-like phase (here the metastable A-like phase is not seen in a whole temperature range) up to temperatures above $T_{ca}$. We interpret the clear resonance features on curves in Fig.2 as signals of longitudinal resonance in the A-like phase (blue curves) and in the B-like phase (red curves). We have found that for a given excitation frequency the maximal response in the A-like phase corresponds to $T\sim 0.835 T_{ca}$ at both used pressures. The corresponding values of $T_{ca}/T_c$ and $T_{ab}/T_{ca}$ are equal to 0.79, 0.81 (for P=26.0 bar) and 0.81, 0.74 (for P=29.3 bar) respectively, where $T_c$ is the superfluid transition temperature of the bulk $^3$He.

It is worth mentioning that further warming up close to $T_c$ results in one more small longitudinal resonance signal which is obviously connected with the longitudinal resonance in the bulk superfluid $^3$He in the gaps between the aerogel and the inner walls of the cell.

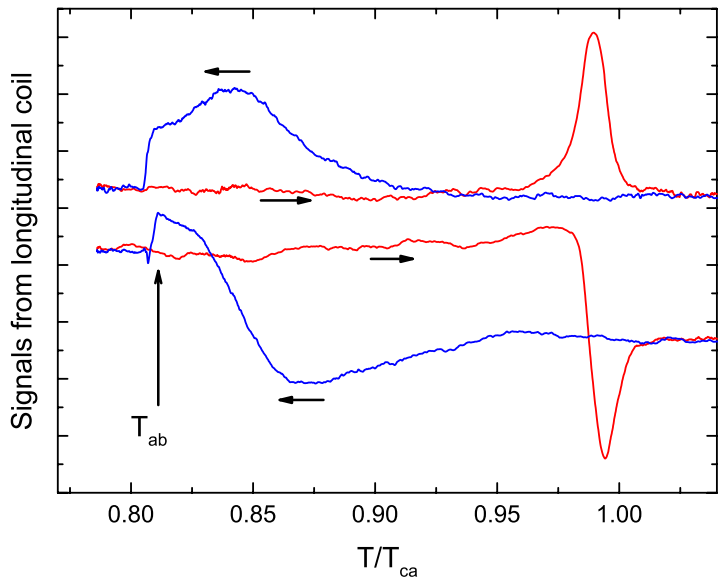

**FIGURE 2.** Longitudinal NMR absorption and dispersion signals versus the temperature. (the B-like phase signal was used as a reference to adjust the lock-in phase angle). Arrows show the direction of the temperature sweep. P=26.0 bar. Note that we extrapolate the mean value of the frequency shift in transverse NMR line to zero to define $T_{ca}$ [13]. This explains why there is still some B-like phase signal in a small region above $T_{ca}$. We can also introduce the temperature $T^*$ above which $^3$He in the whole volume of aerogel is in normal state (usually $T^*=T_{ca}(1+k)$ and $k\sim 0.01$).

## TRANSVERSE NMR EXPERIMENTS

It was found that if the A-like phase is obtained as it is described above (i.e. by cooling from the normal phase without additional external perturbations) then the continuous wave NMR line has a two-maximum shape (Fig. 3). These two distinct maximums can be attributed to some complex texture of the order parameter. Another possibility (which is more probable as it will be seen below) is that we have two different textures (two different spin states) inside the sample. Each of these states corresponds to some well defined value of the NMR frequency shift. CW NMR experiments in presence of longitudinal gradient of the external steady magnetic field have shown that these states are not homogeneously distributed over the sample but concentrated mainly at the opposite ends of the sample: the state with the larger frequency shift (let us call it "far from the Larmor value" or "f"-state) is situated mostly in the lower part of the sample and the state with the smaller shift ("close to the Larmor value" or "c"-state) – in the upper part. Let us suggest that the NMR line in each state can be described by a simple Lorentzian. Then we obtain (by a simple fit with two Lorentzians) that in a wide range of temperatures the ratio of the frequency shifts in "f" and "c" states is about 4. Volumes occupied by these states are usually equal (for 29.3 bar the ratio of integrals over the "f"-state Lorentzian and "c"-state Lorentzian were 1.03 ±0.05 in a number of runs; at 26.0 bar sometimes we obtained 2-3 times less amount of the "f"-state).

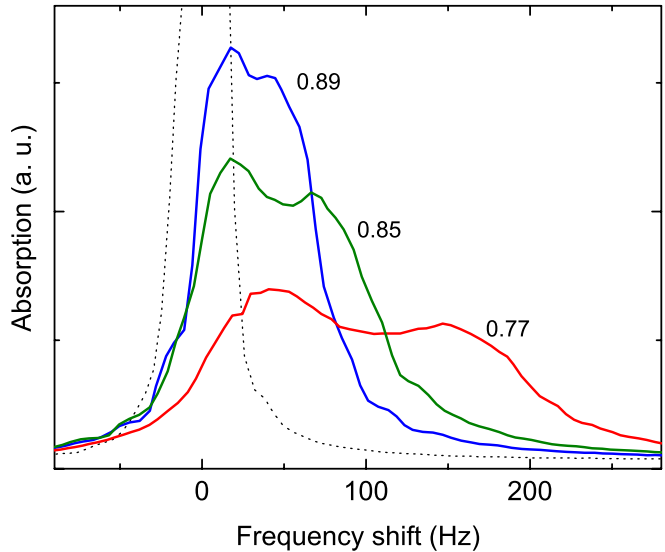

**FIGURE 3.** CW NMR absorption signals in the A-like phase at different temperatures. Temperatures in terms of $T/T_{ca}$ are shown near each curve. Dotted curve corresponds to $T>T_{ca}$. P=29.3 bar.

In order to get pure ("f" or "c") state we have tried to obtain the A-like phase by fast (or very slow) cooling down through $T_{ca}$. However the two-maximum shaped line was found to be reproducible. Therefore we have tried another technique to influence the texture, i.e. the application of 180 degree tipping pulses. It was shown earlier that in the bulk A-phase such a method results in essential changes in the shape of the NMR line due to creation of solitons [14,15]. We have found that in our case the application of 180 degree tipping pulses well below $T_{ca}$ does not influence the shape of the CW NMR line. However, if we apply 180 degree pulses while cooling down through $T_{ca}$ then the situation is quite different. The "f"-state disappears and CW NMR line corresponding to nearly pure "c"-state is observed (Fig.4). The above mentioned procedure of fitting the NMR line by two lorentzians gives a good agreement between "c"-component of the mixed "f+c"-state and pure "c"-line (Fig.5). Even more remarkable is the following observation: the longitudinal NMR signal from the pure "c"-state is absent. This explains also why the longitudinal signal from the mixed "f+c"-state (blue curves in Fig.2) is not a two-maximum shaped line: we see in fact only the signal from the "f"-state.

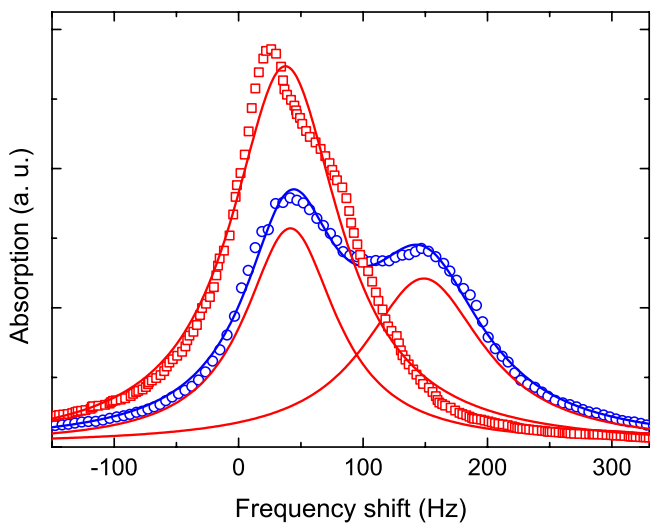

**FIGURE 4.** CW NMR lines in the A-like phase at the same experimental conditions. Blue circles – "f+c"-state, red squares – pure "c"-state (obtained after application of 180 degree pulses at temperatures near $T_{ca}$). Solid lines are the best fit by Lorentzians. P=29.3 bar, T=0.77 $T_{ca}$.

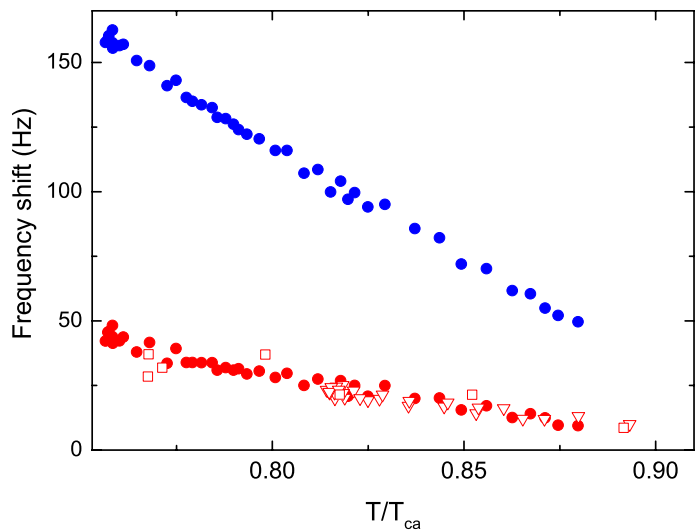

**FIGURE 5.** Frequency shifts obtained by fitting CW NMR lines with Lorentzians. Blue and red circles: "f"-state and "c"-state components in the mixed "f+c"-line respectively; open squares and triangles: pure "c"-state (obtained at different cool downs). P=29.3 bar.

## PULSED NMR EXPERIMENTS

In both pure "c" and mixed ("f+c") states we have measured the dependence of the frequency $\omega$ of free induction decay signal (FIDS) versus the initial tipping angle $\beta$. The frequency was determined by fitting the initial ~1 millisecond of the observed FIDS (which decay time was about 10 ms) by a function $a(t)\sin(\omega t+\varphi)$. Calibrations in the normal phase have shown that for $\beta<140$ degrees the error in determination of the frequency does not exceed 3 Hz and is defined mainly by the stability of the steady magnetic field. For $\beta>140$ degrees the results are not reliable: the fitted frequency of the FIDS even in the normal phase starts to depend on $\beta$. We think that it is due to the additional signal(s) from the regions outside the aerogel cell. Therefore below we present only points up to $\beta=141$ degrees where the estimated error is not yet large (~10 Hz).

The dependence shown in Fig.6 is obtained in pure "c"-state and it can be fitted by $\omega-\omega_L=A\cos\beta$, where $\omega_L$ is the Larmor frequency. The results for the mixed "f+c" state are shown in Fig.7. Note that there is an additional complication in this case. The effects of fast spin exchange in the mixed state are presumably not large (otherwise we would not observe two-maximum shape of the CW NMR line or at least the width and position of the pure "c"-line would not correspond to the width and position of the "c"-component of the mixed state line). Correspondingly, the FIDS in the "f+c"-state should be a sum of two FIDS (from "f" and "c" states) and our procedure of fitting by one sinusoid does not describe the situation properly: the value of the fitted frequency may depend on relative amplitudes of FID signals from "f" and "c" states which may change with the time in a different way, on frequencies of these states (which also may depend on time) etc. Our attempts to fit FIDS in the mixed state with a sum of two sinusoids were not a success yet. Thus, the results presented in Fig.7 may be considered only as a first approximation, if one wants to compare them with the theoretical predictions. It is worth mentioning that the results shown in Fig.7 qualitatively agree with pulsed NMR results in 97.5% aerogel [6].

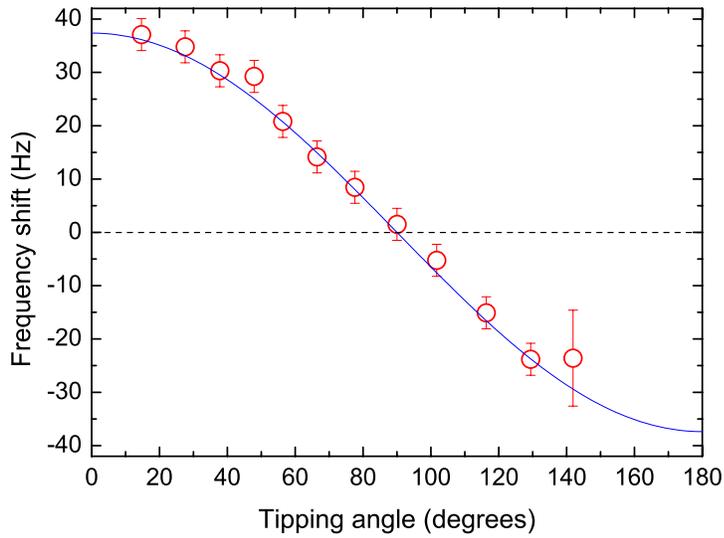

**FIGURE 6.** FIDS frequency versus the initial tipping angle in pure "c"-state. Solid line is the best fit by $(\omega-\omega_L)/2\pi=A\cos\beta$ with A=37.4 Hz. P=29.3 bar, T=0.76$T_{ca}$. It agrees with the shift of the CW line in the "c"-state under the same conditions (Fig.5).

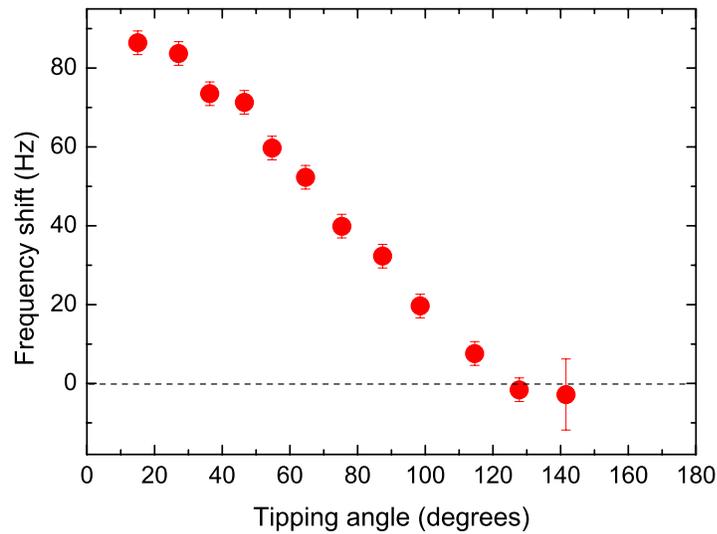

**FIGURE 7.** FIDS frequency versus the initial tipping angle in the mixed "f+c"-state. P=29.3 bar, T=0.76$T_{ca}$.

**DISCUSSION**

Our results are summarized below:

1. Longitudinal NMR signals have been observed in both A-like and B-like superfluid phases.

2. When the A-like phase is obtained on cooling without additional effort we get a two-maximum shaped CW absorption signal ("f+c"-state). If at temperatures near $T_{ca}$ we apply 180 degrees pulses, then CW NMR line is changed: only one of the above mentioned maximums (with a smaller frequency shift) remains. We call this state "c". Remarkable feature of the "c"-state is that we see no traces of the longitudinal resonance. It means that either the longitudinal resonance in this state is absent, or its frequency in our temperature range (0.74 $T_{ca}$ – 1.0 $T_{ca}$) is much smaller than 9.593 kHz.

3. The frequency of FIDS from the "c"-state is well described by the following dependence:

$(\omega-\omega_L)/2\pi = A(T)\cos\beta$

The origin of the "c" state (as well as the dependence of the FIDS frequency in the "c"-state) can be explained by the theory of the "nearly robust" ESP phase [16], where it is shown that this state can be a metastable spin-orbital configuration of the "robust" ESP phase. However many unclear points still remain. At P=29.3 bar transverse NMR frequency shift in the "f"-state at T~0.835$T_{ca}$ (where we see the maximal absorption in longitudinal resonance) is about 85 Hz (blue circles in Fig.5). Let us suggest that the "f"-state is purely "robust" ESP phase. Using the results of [17] we can find that for a given longitudinal resonance frequency we should have the shift about 32 Hz. Theory developed in [10] gives approximately the same results as [17] for the relation between transverse NMR frequency shift and the frequency of the longitudinal resonance. Corrections due to possible deflections from the "robust" state [16] can improve the situation, but not sufficiently. A good agreement with [16] can be obtained only if we suggest that for some reason the measured value of the longitudinal resonance is 2/3 of the Leggett frequency in the A-like phase. It is possible if nevertheless there is some interaction between "f" and "c" states, however this interaction also should change the results for the transverse NMR shift.

It is also not clear why in some experiments in the A-like phase a negative frequency shift from the Larmor value is observed [1,4,5]. In one of our previous cells the maximal negative frequency shift was about 600 Hz (in field of 284 Oe and at pressure of 25.5 bar) [5]. Note that now (at similar conditions of the experiments) the shift is positive and does not exceed ~200 Hz. In another sample of aerogel (the cell is described in [13]) we have seen only positive frequency shift and its value (recalculated to 224 Oe) was close to the case of our present aerogel sample. One more question is why the frequency of longitudinal magnetic resonance in the A-like phase is much smaller than in the B-like phase. It is not clear why "f"-state does not survive after 180 degree pulses. Further experiments are necessary to clarify these questions. In particular, it would be quite useful to obtain pure "f"-state and to measure the dependence of its FIDS frequency on the tipping angle and to observe the longitudinal resonance.

## ACKNOWLEDGEMENTS

We thank I.A.Fomin and G.E.Volovik for stimulating discussions and I.A.Fomin for the suggestion to try 180 degree pulses. This work was supported by CRDF (grant #RUP1-2632-MO-04), RFBR (grant 06-02-17185) and by the Ministry of Education and Science of Russia. D.E.Z. thanks the Russian Science Support Foundation and the Forschungszentrum Juelich for financial support within the framework of the Landau Program.